\documentclass[english,12pt,a4paper]{article}
\usepackage{authblk}
\usepackage[text={17.0cm,23.7cm}]{geometry}
\usepackage[T1]{fontenc}
\usepackage[utf8]{inputenc}
\usepackage[english]{babel} 
\usepackage{hyperref}
\usepackage{graphicx}
\usepackage{amsmath}
\usepackage{amssymb}
\usepackage[dvipsnames]{xcolor}
\usepackage{yfonts}
\usepackage[mathscr]{euscript}
\usepackage{float}
\usepackage{csquotes}
\usepackage[normalem]{ulem} 
\usepackage{verbatim}
\usepackage[backend=biber,style=numeric-comp,sorting=none]{biblatex}

\usepackage{bm} 
\usepackage{bbold} 
\usepackage{relsize} 
\usepackage{caption}
\usepackage{subcaption} 
\usepackage{pifont} 

	\title{\vspace{-2cm}
		\vspace{0.6cm}
	\textbf{Cosmological signals of dark matter semi-annihilation}\\[8mm]}
    
\begin{document}
\author[1]{Boris Betancourt Kamenetskaia}
\author[2]{Mathias Garny}
\author[2]{Alejandro Ibarra}
\author[2]{Alessia Musumeci}
\author[3]{Merlin Reichard}

\affil[1]{\normalsize\textit{Cosmology, Gravity, and Astroparticle Physics Group, Center for Theoretical Physics of
the Universe, Institute for Basic Science (IBS), Daejeon, 34126, Korea}}
\affil[2]{\normalsize\textit{Technical University of Munich, TUM School of Natural Sciences, Physics Department, 85748 Garching, Germany}}

\affil[3]{\normalsize\textit{School of Physics, Korea Institute for Advanced Study, Seoul 02455, Republic of Korea}}

\date{}
\maketitle

\begin{abstract}

The growth of primordial density fluctuations in the early Universe leads to an inhomogeneous dark matter distribution at high redshift, where semi-annihilation processes of the form $\chi\chi \rightarrow \chi^c \phi$, with $\phi$ being dark radiation, can occur with a sizable rate. Using a state-of-the-art model for the cosmological boost factor, we compute the resulting redshift-dependent flux of boosted dark matter particles generated by semi-annihilation, and we study the implications of the boosted component for structure formation and direct detection experiments. We find a model independent upper limit on the semi-annihilation cross-section from structure formation, which reads $
\langle\sigma_{2\to1} v\rangle\leq4.2\times10^{-19}~\left(m_\chi/1~\rm GeV\right)~\mathrm{cm}^3/{\rm s}$. Further, we find that the cosmological contribution to the boosted dark matter flux can be comparable to the galactic one, providing an $O(1)$ enhancement to the sensitivity of dark matter searches, thus slightly enhancing the discovery potential in direct detection experiments of semi-annihilation scenarios where the dark matter interacts with the nucleus. 
\end{abstract}
\section{Introduction}
\label{sec:Intro}

Two necessary ingredients of any viable dark matter (DM) model are a mechanism that generates a population of DM particles in the very early Universe, as well as a mechanism that guarantees their stability on cosmological timescales. Many DM scenarios feature a discrete $Z_2$ symmetry (either accidental or imposed), under which the DM fields are odd while the Standard Model fields are even (notable examples are R-parity in supersymmetric scenarios~\cite{Farrar:1978xj,Goldberg:1983nd,Ellis:1983ew} or the Kaluza–Klein parity symmetry in scenarios with universal extra dimensions~\cite{Appelquist:2000nn,Servant:2002aq}). This symmetry ensures the stability of DM, allows DM particles to thermalize with the Standard Model plasma, and leads to the annihilation of two DM particles into two Standard Model particles. These annihilations continue until the expansion rate of the Universe exceeds the annihilation rate, leaving behind a relic abundance of DM particles today.

In another class of scenarios, the DM field transforms under a discrete $Z_3$ symmetry~\cite{Hambye:2008bq,Hambye:2009fg,Arina:2009uq,DEramo:2010keq,Belanger:2012zr,Belanger:2014bga,Aoki:2014cja,Bandyopadhyay:2022tsf}, which also guarantees DM stability and allows the thermalization of the DM with the Standard Model. However, in this case pair annihilations into two Standard Model particles are forbidden by the symmetry. On the other hand, processes in which two DM particles annihilate into one DM particle and one Standard Model particle are allowed. This process, dubbed ``semi-annihilation,'' also depletes the comoving DM density in the early Universe and can generate a relic abundance of DM particles, similarly to the more commonly studied annihilation scenario.

One of the most salient features of the semi-annihilation scenario is the generation of a boosted DM (BDM) component, since part of the energy of the initial state is converted into kinetic energy of the DM particle in the final state. This boosted component enhances the detection prospects of DM in direct detection experiments, especially for masses in the sub-GeV range, where the halo component does not induce detectable nuclear recoils~\cite{Bringmann:2018cvk,Dent:2019krz,Bondarenko:2019vrb}. Indeed, if semi-annihilations take place in the Milky Way center with  $\langle \sigma_{2\to1} v\rangle \sim 10^{-26}\,{\rm cm}^3/{\rm s}$, which is in the ballpark of the values required for production via thermal freeze-out, the XENONnT experiment excludes DM-proton scattering cross sections as low as $10^{-36}\,{\rm cm}^2$ for a DM mass of $\sim 30$ MeV, while the CRESST-II experiment excludes  cross sections of order $\sim 10^{-31}\,{\rm cm}^2$ for DM masses around $\sim 10$ MeV~\cite{BetancourtKamenetskaia:2025noa}. 

Semi-annihilations are expected to occur not just in the Milky Way center but in any region with a DM overdensity. In this work, we focus on semi-annihilations taking place in cosmological structures 
and we investigate their possible signatures~\cite{Bergstrom:2001jj,Ullio:2002pj}. In Section~\ref{sec:flux}, we calculate the BDM flux and study the implications of the warm DM component injected into the Universe for structure formation. Then, in Section~\ref{sec:DD}, we investigate the implications for direct detection experiments of the BDM flux reaching the Earth, assuming that DM interacts with nucleons. Finally, in Section~\ref{sec:conclusions}, we present our conclusions.

\section{Boosted dark matter flux from semi-annihilation}
\label{sec:flux}

We consider the DM semi-annihilation process
$\chi \chi \to \chi^c \phi$ occurring in DM clumps at
cosmological redshift with cross section times velocity $\langle\sigma_{2\to1} v\rangle$. The semi-annihilation at the redshift $z$ produces a BDM particle with kinetic energy $T_z=m_\chi/4$. The number of BDM particles produced per unit volume, unit time and unit kinetic energy is given by:
\begin{equation}\label{eq:Q}
Q_{\rm BDM}(T_z,z)=
\frac{1}{2}\,\langle\sigma_{2\to1} v\rangle
\frac{\rho_\chi^2(z)}{m_\chi^2}\,
G(z)\,\delta\left(T_z-\frac{m_\chi}{4}\right),
\end{equation}
where 
$\rho_\chi(z)=\Omega_{\chi,0}\rho_{c,0}(1+z)^3$ is the redshift-dependent cosmological DM density, expressed in terms of the critical density today $\rho_{c,0}\simeq 5\times 10^{-6}\,{\rm GeV}/{\rm cm}^3$ and the DM density 
parameter today $\Omega_{\chi,0}\simeq 0.26$~\cite{Planck:2018vyg}, while 
$G(z)$ parametrizes the enhancement of the annihilation rate due to substructures, for which we adopt the value calculated in Ref.~\cite{Lopez-Honorez:2013cua}.\footnote{We checked that cutting off the halo mass function at $M\sim 10^{10} (10^{3}) M_\odot$ reduces the boost factor at redshifts $z\lesssim 10$ by at most a factor of around $2$ ($5$). Thus, even if the halo mass function differs from the one applicable to cold dark matter on small scales, our conclusions are robust at the level quoted above.} Then, the total BDM flux at redshift $z$ is obtained from integrating the source term over the cosmological line-of-sight and over all incoming directions:
\begin{equation}\label{eq:diff_flux}
    \frac{d\Phi_{\rm BDM}}{dT_z}(T_z,z) = (1+z)^2
\int_z^\infty
\frac{dz'}{H(z')(1+z')^3} \sqrt{1-\left(\frac{m_\chi}{T_{z'}+m_\chi}\right)^2}Q_{\rm BDM}\left(T_{z'},z'\right),
\end{equation}
where $H(z)= H_0 \, \sqrt{\Omega_{\rm m,0}(1+z)^{3} +  \Omega_{\Lambda}}$ is the Hubble parameter as a function of redshift, with $\Omega_{\rm m,0}=1-\Omega_\Lambda\simeq 0.31$ and $H_0\simeq 67\,{\rm km}\,{\rm s}^{-1}\,{\rm Mpc}^{-1}$~\cite{Planck:2018vyg}.
Further, $T_{z'}$ is the kinetic energy of the BDM particle at the redshift $z'$, which is related to the kinetic energy at a different redshift $z$ by:
\begin{align}
T_{z'}=\sqrt{m_\chi^2+ \left(\frac{1+z'}{1+z}\right)^2
T_{z}(T_{z}+2m_\chi)}- m_\chi.
\end{align}

Replacing the source term and integrating one obtains
\begin{equation}\label{eq:diff_flux_general}
    \frac{d\Phi_{\rm BDM}}{dT_z}(T_z,z) = \frac{2(1+z)^2}{3}\,\frac{\rho_\chi^2(z_*)}{m_\chi^2}\,\frac{\langle\sigma_{2\to1} v\rangle G(z_*)}{(1+z_*)^2H(z_*)m_\chi}\Theta\left(z_*-z\right),
\end{equation}
with $z_*$ determined from
\begin{equation}
    \frac{(1+z_*)}{ (1+z)} = \frac{3}{4}\left[\frac{T_z}{m_\chi}\left(\frac{T_z}{m_\chi}+2\right)\right]^{-\frac12}.
\end{equation}

\begin{figure}[t!]
\centering
\includegraphics[width=0.58\textwidth]{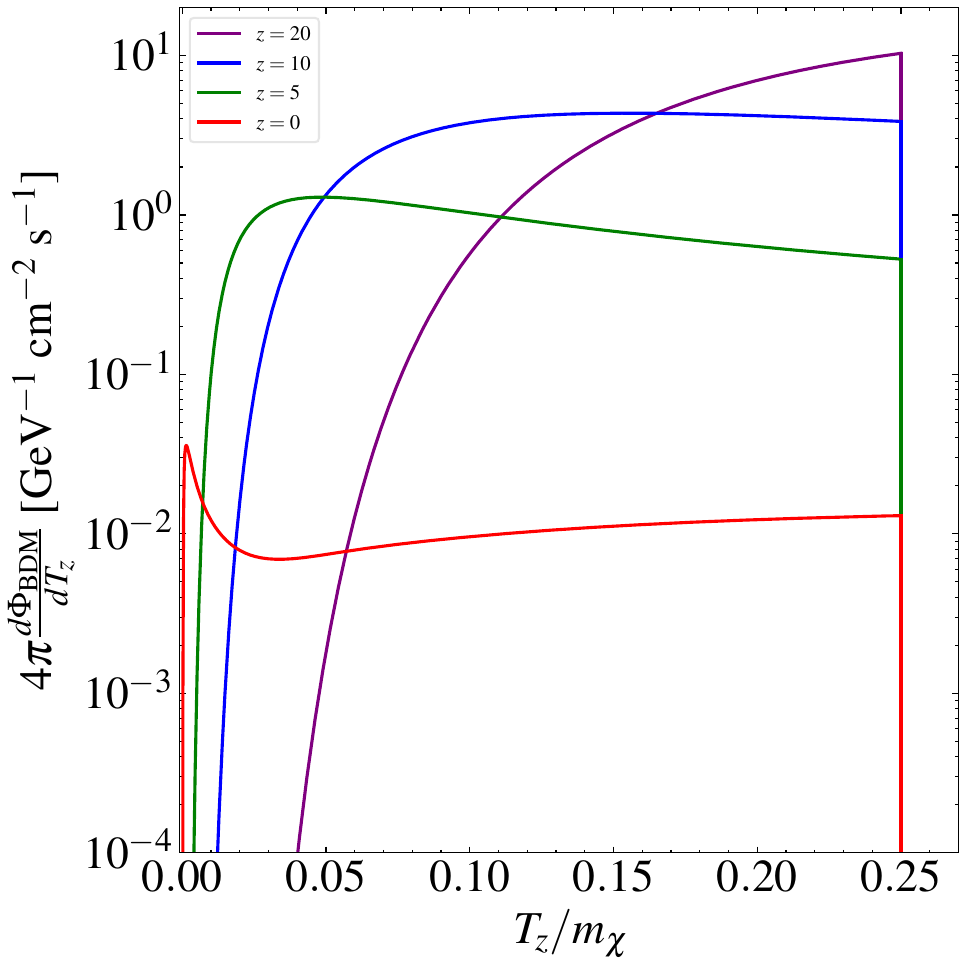}
\caption{Differential flux of BDM particles from semi-annihilation at different redshifts, assuming for concreteness $m_\chi=100$ MeV and $\langle\sigma_{2\to1} v\rangle=10^{-26}~\mathrm{cm}^3~\mathrm{s}^{-1}$.}
\label{fig:DiffFluxes}
\end{figure}

In Fig.~\ref{fig:DiffFluxes}, we present the differential fluxes for a semi-annihilation process as a function of $T_z/m_\chi$ at different redshifts. For concreteness, we choose $m_\chi=100~\mathrm{MeV}$ and $\langle\sigma_{2\to1} v\rangle=10^{-26}~\mathrm{cm}^3~\mathrm{s}^{-1}$; for other values of the mass and cross section, the vertical axis should be correspondingly scaled by a factor $(\langle \sigma_{2\to1} v\rangle/10^{-26}\,{\rm cm}^3\,{\rm s}^{-1}) (m_\chi/100\,{\rm MeV})^{-3}$.  It is notable from the plot that the flux of BDM from cosmological semi-annihilations can be sizable, even of the order of a few particles per ${\rm cm}^2 \,{\rm s}$. Further, one finds that the peak of the energy spectrum moves towards lower values for decreasing $z$ and the flux at large redshift is larger than at low redshift. Both are a consequence of the effect of the redshift on the energy of the BDM particles, combined with the non-trivial dependence of the cosmological boost factor with the redshift, as well as the ``stretching'' of the distance between the source and the Earth due to the expansion of the Universe.

In particular, we note that at the range of redshifts where the formation of the first baryonic structures takes place, there could be a significant number of DM particles with a large free-streaming length, thereby erasing small scale structures. To the best of our knowledge, there is no dedicated analysis evaluating the maximally allowed fraction of warm DM generated in cold DM (semi-)annihilations. Therefore, we will adopt the conservative upper limit $f_{\rm BDM}\leq 0.01$ at $z=0$, which roughly corresponds to the 95\% C.L. upper limit $f_{\rm DCDM}\leq 0.022$ derived from the BOSS-DR12 data~\cite{BOSS:2016wmc} analyzed under the Effective Field Theory approach to large-scale structure~\cite{Simon:2022ftd}.\footnote{The precise limit depends on the redshifts at which the warm DM population is produced. In order to obtain an estimate we approximately recast limits derived for decaying dark matter. To bracket the uncertainty from the different redshift distribution of the semi-annihilation rate as compared to the decay rate, we consider two limiting cases, one in which the production occurs dominantly at high redshift, and one at which it is dominated by low redshifts. The former corresponds to the scenario in which a fraction of cold dark matter decays with a lifetime much shorter than the age of the Universe, for which a combination of Planck CMB and BOSS-DR12 data yields $f_{\rm DCDM}\leq 0.022$ at 95\% C.L. as quoted above. For the latter, we consider the case in which all of DM is assumed to decay eventually. The same data sets then yield the limit $\tau>250$\,Gyr on the lifetime at 95\% C.L.~\cite{Simon:2022ftd}. The fraction of DM that has decayed at time $t$ is $f_{\rm DCDM}(t)=1-e^{-t/\tau}\simeq t/\tau$, which yields $f_{\rm DCDM}\leq 0.055$ when evaluated at the present time. The redshift-dependence of the semi-annihilation rate is in between these two scenarios, such that the upper limits for DCDM for the two scenarios may be expected to provide an estimate of the range of the BDM fraction allowed by CMB and large-scale structure data. The limit adopted in this work can thus be viewed as conservative, while a more precise determination would require a dedicated analysis.}

\begin{figure}[t!]
\centering
\includegraphics[width=0.48\textwidth]{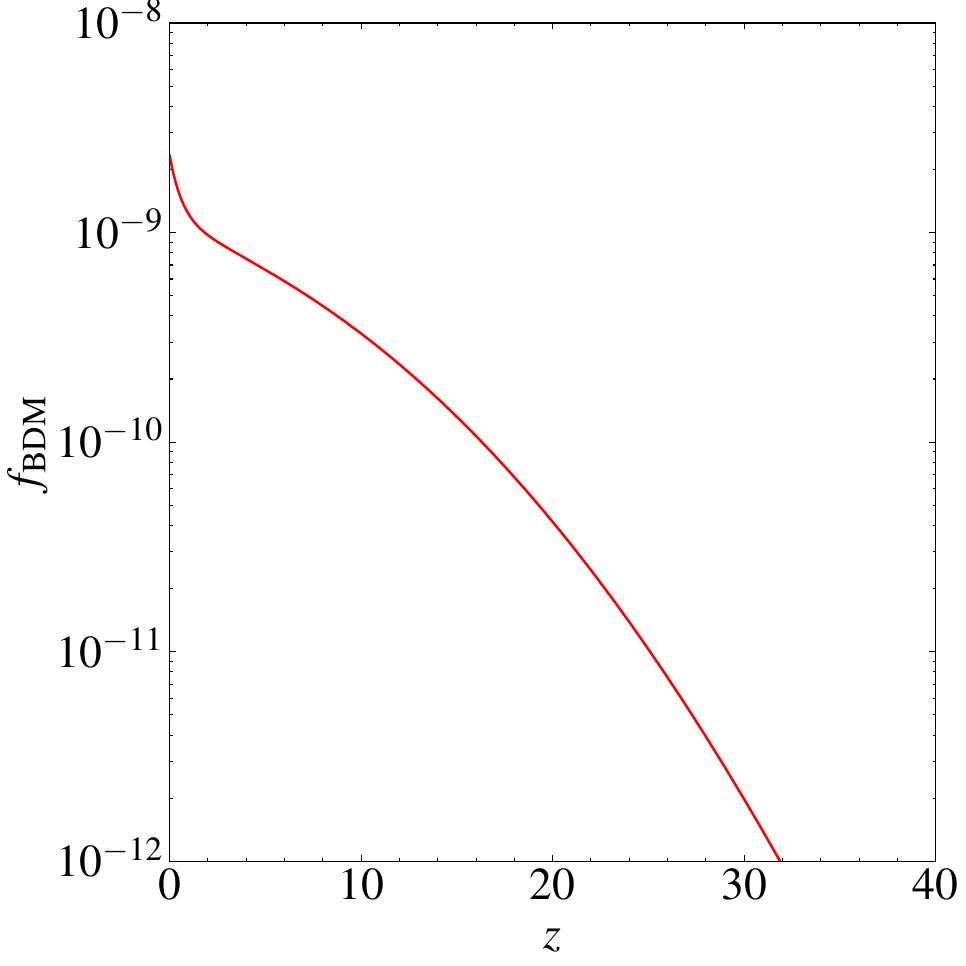}
\includegraphics[width=0.48\textwidth]{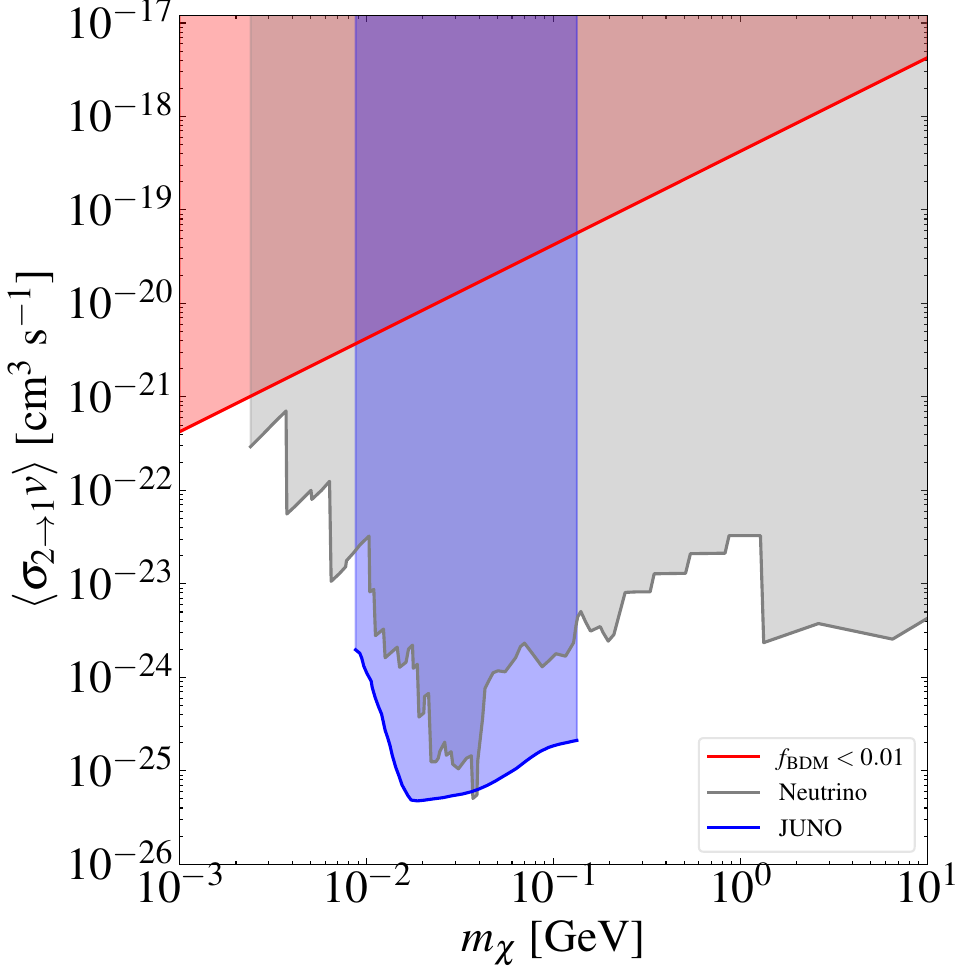}
\caption{{\it Left panel}: Fraction of BDM as a function of the redshift, for $m_\chi=100~\rm MeV$ and $\langle\sigma_{2\to1} v\rangle=10^{-26}~\mathrm{cm}^3~\mathrm{s}^{-1}$. {\it Right panel}: Constraints on $\langle\sigma_{2\to1} v\rangle$ as a function of DM mass from structure formation (red area) and from neutrino telescopes (grey area), along with the projected JUNO sensitivity.}
\label{fig:BDM_constraints}
\end{figure}

The number density of BDM particles produced in semi-annihilations at redshift $z$ can be obtained from the flux: 
\begin{equation}\label{eq:n_BDM}
    n_{\rm BDM}(z)=\int \frac{1}{v(T_z)}\frac{d\Phi_{\rm BDM}}{dT_{z}}dT_{z}=\frac{\langle\sigma_{2\to1} v\rangle(1+z)^3}{2m_\chi^2}\int\displaylimits_{z}^{\infty} dz' \frac{\rho_\chi^2(z') G(z')}{(1+z')^4 H(z')},
\end{equation}
with 
\begin{equation}
    v(T_z)=\sqrt{1-\left(\frac{m_\chi}{T_z+m_\chi}\right)^2}.
\end{equation}
From Eq.~(\ref{eq:n_BDM}) it is straightforward to calculate the fraction of BDM particles, defined as
\begin{align}
f_{\rm BDM}(z)\equiv \frac{n_{\rm BDM} (z)}{n_{\chi}(z)},
\end{align}
with $n_{\chi}(z)=\rho_{\chi}(z)/m_\chi$. The fraction as a function of redshift is shown in Fig.~\ref{fig:BDM_constraints}, left panel, for the specific case  $m_\chi=100~\rm MeV$ and $\langle\sigma_{2\to1} v\rangle=10^{-26}~\mathrm{cm}^3~\mathrm{s}^{-1}$. Finally, imposing $f_{\rm BDM}(z=0)\leq 0.01$ we obtain
\begin{align}
\langle\sigma_{2\to1} v\rangle\leq4.2\times10^{-19}~\left(\frac{m_\chi}{1~\rm GeV}\right)~\mathrm{cm}^3/{\rm s}.
\end{align}

The excluded region is shown as the red-filled area in the right panel of Fig.~\ref{fig:BDM_constraints}. We note that if the dark radiation is identified with neutrinos, the neutrino flux at Earth is expected to receive contributions from these semi-annihilation events occurring in the Galaxy and from extragalactic origins. The regions of parameter space excluded by the non-observation of an excess in the neutrino flux from the semi-annihilation $\chi\chi\rightarrow\chi^c \nu$ are also shown in the Figure, obtained from adapting the constraints from~\cite{Arguelles:2019ouk} for the annihilation $\chi\chi\rightarrow \bar \nu \nu$, to the corresponding neutrino energy and to the number of neutrinos in the final state. Finally, the blue region shows the sensitivity reach of the JUNO experiment, as calculated in Ref.~\cite{Akita:2022lit}. It follows from the plot that the structure formation constraints are weaker than those from neutrino observations, except for $m_\chi\lesssim 2$ MeV, where the neutrino does not generate a detectable signal in neutrino telescopes. 

If the DM particle has additional interactions with the Standard Model aside from the neutrino, the flux of BDM particles from cosmological semi-annihilations could produce signatures in direct detection experiments. These are discussed in the next section.

\section{Implications of the boosted dark matter flux for direct detection experiments}
\label{sec:DD}

The differential event rate per unit target mass for the scattering of a DM particle off a nucleus ${\cal T}$, inducing a nuclear recoil with kinetic energy $T_{\cal T}$, is given by
\begin{equation}
    \frac{dR_{\cal T}}{dT_{\cal T}}
    =
    \frac{1}{m_{\cal T}}
    \int\displaylimits_{T_{\chi,{\cal T}}^{\rm min}(T_{\cal T})}^{\infty}
    dT_\chi\,
    \frac{d\sigma_{\chi {\cal T}}}{dT_{\cal T}}(T_\chi,T_{\cal T})
    \frac{d\Phi_{\rm BDM}}{dT_\chi}\, .
    \label{eq:diff_rate}
\end{equation}
Here, $m_{\cal T}$ denotes the mass of the target nucleus, while $T_\chi$ is the kinetic energy of the incoming DM particle. The differential flux $d\Phi_{\rm BDM}/dT_\chi$ is understood as that given in Eq.~\eqref{eq:diff_flux_general} with $z=0$. The quantity $T_{\chi,{\cal T}}^{\rm min}$ corresponds to the minimum kinetic energy required for the incoming DM particle to produce a recoil energy $T_{\cal T}$, and is given by
\begin{equation}
\label{eq:T_min}
    T_{\chi,{\cal T}}^{\rm min}(T_{\cal T})
    =
    \left(
    \frac{T_{\cal T}}{2}-m_\chi
    \right)
    \left[
    1\pm
    \sqrt{
    1+
    \frac{2T_{\cal T}}{m_{\cal T}}
    \frac{(m_\chi+m_{\cal T})^2}{(T_{\cal T}-2m_\chi)^2}
    }
    \right],
\end{equation}
where the plus (minus) sign applies for $T_{\cal T}>2m_\chi$ ($T_{\cal T}<2m_\chi$).

Further, $d\sigma_{\chi{\cal T}}/dT_{\cal T}$ denotes the differential scattering cross section for the interaction of DM with the nucleus ${\cal T}$. Since DM particles produced through semi-annihilations are boosted (see Eq.~(\ref{eq:Q})), the momentum transfer can become sufficiently large that its inverse is comparable to, or smaller than, the nuclear size. In this regime the scattering gradually loses coherence over the entire nucleus. Following Refs.~\cite{Bednyakov:2018mjd,Bednyakov:2021bty}, we parametrize this effect by separating the total differential cross section into coherent and incoherent contributions:
\begin{align}
 \frac{d\sigma_{\chi{\cal T}}}{dT_{\cal T}}
 =
 \left(
 \frac{d\sigma_{\chi{\cal T}}}{dT_{\cal T}}
 \right)_{\rm coh}
 +
 \left(
 \frac{d\sigma_{\chi{\cal T}}}{dT_{\cal T}}
 \right)_{\rm inc}\, ,
 \label{eq:diff_cross_section}
\end{align}
which smoothly interpolates between the coherent and incoherent scattering regimes. 
For spin-independent interactions, these contributions read
\begin{align}
\label{eq:diff_scat_sigma}
    \left(
    \frac{d\sigma_{\chi{\cal T}}}{dT_{\cal T}}
    \right)_{\rm coh}
    &=
    \frac{\sigma^{\rm coh}_{{\rm SI},{\cal T}}}
    {T_{\cal T}^{\rm max}}
    |F_{{\rm SI},{\cal T}}(q)|^2,
    \\
    \left(
    \frac{d\sigma_{\chi{\cal T}}}{dT_{\cal T}}
    \right)_{\rm inc}
    &=
    \frac{\sigma^{\rm inc}_{{\rm SI},{\cal T}}}
    {T_{\cal T}^{\rm max}}
    \left(
    1-|F_{{\rm SI},{\cal T}}(q)|^2
    \right)\,,
\end{align}
where
\begin{equation}
T_{\cal T}^{\rm max} = \frac{T_\chi^2+2m_\chi T_\chi}
{T_\chi+(m_\chi+m_{\cal T})^2/(2m_{\cal T})},
\label{eq:T_max}
\end{equation}
is the maximum recoil energy that can be transferred to the target nucleus, while
$q=\sqrt{2m_{\cal T}T_{\cal T}}$
is the momentum transfer.

For the nuclear form factor we adopt the dipole approximation~\cite{Perdrisat:2006hj},
\begin{equation}
\label{eq:form_factor}
    F_{{\rm SI},{\cal T}}(q)
    =
    \left(
    1+\frac{q^2}{\Lambda_{\cal T}^2}
    \right)^{-2},
\end{equation}
with
$\Lambda_{\cal T}\simeq 0.843\, \mathrm{GeV}
\left(
0.8791\,{\rm fm}/R_{\cal T}
\right)$,
where $R_{\cal T}$ is the charge radius of the target nucleus (see e.g. Ref.~\cite{Angeli:2004kvy} for a compilation of charge radii).

The quantities
$\sigma^{\rm coh}_{{\rm SI},{\cal T}}$
and
$\sigma^{\rm inc}_{{\rm SI},{\cal T}}$
correspond respectively to the coherent and incoherent spin-independent scattering cross sections at zero momentum transfer. They are related to the DM-proton and DM-neutron cross sections, $\sigma_p$ and $\sigma_n$, through
\begin{align}
    \sigma^{\rm coh}_{{\rm SI},{\cal T}}
    &=
   \sigma_p
   \left(
   \frac{\mu_{\cal T}}{\mu_p}
   \right)^2
   \left[
   Z_{\cal T} f^p
   +(A_{\cal T}-Z_{\cal T})f^n
   \right]^2,
   \\
    \sigma^{\rm inc}_{{\rm SI},{\cal T}}
    &=
    Z_{\cal T}\sigma_p
    +(A_{\cal T}-Z_{\cal T})\sigma_n\, ,
\end{align}
where
$\mu_{p,n}=m_{p,n} m_\chi/(m_{p,n}+m_\chi)$
is the reduced mass of the DM-nucleon system. Here, $A_{\cal T}$ and $Z_{\cal T}$ denote the mass and atomic numbers of the target nucleus, respectively, while $f^{p,n}$ parametrize the couplings to protons and neutrons.

\begin{figure}[t!]
\centering
\includegraphics[width=0.5\textwidth]{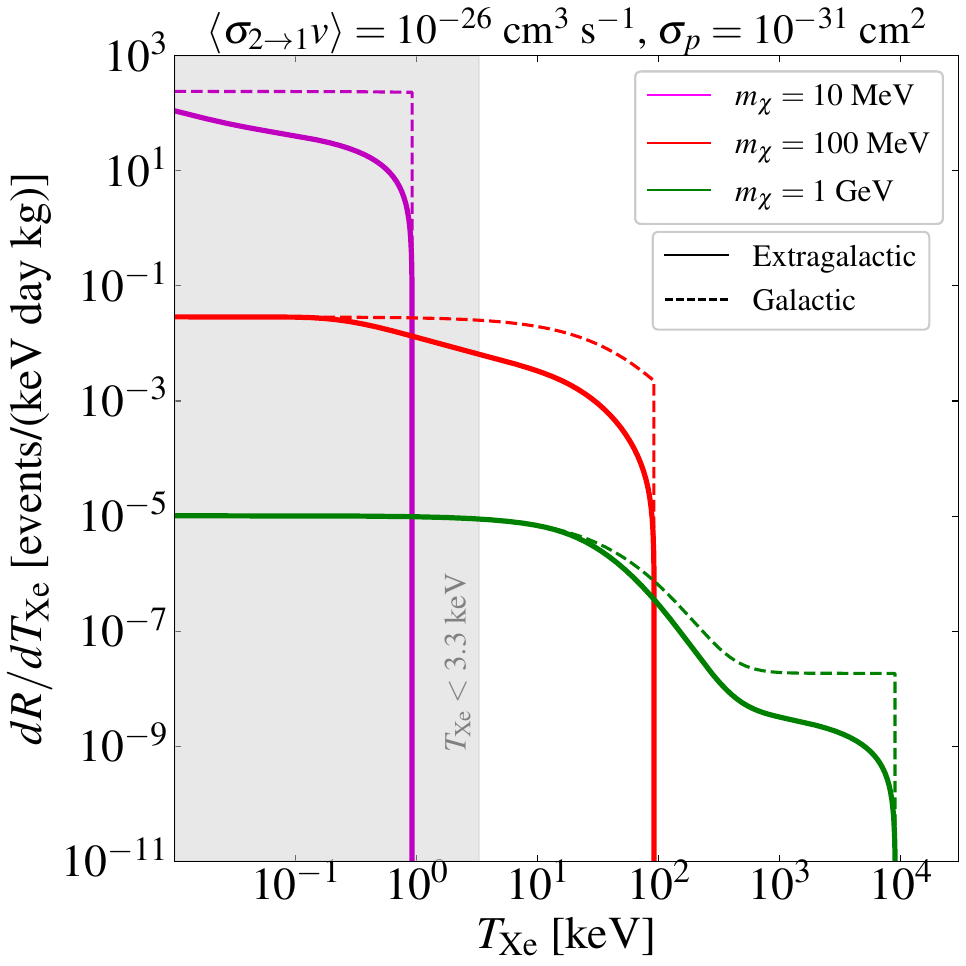}
\caption{Differential recoil rate for DM scattering on a $^{131}$Xe nucleus as a function of the recoil energy. We consider DM masses
$m_\chi=10~\rm MeV$ (violet),
$m_\chi=100~\rm MeV$ (red),
and
$m_\chi=1~\rm GeV$ (green).
The semi-annihilation cross section is fixed to
$\langle\sigma_{2\to1} v\rangle=10^{-26}~\mathrm{cm}^3~\mathrm{s}^{-1}$,
while the DM--nucleon cross section is chosen as
$\sigma_p=10^{-31}\,{\rm cm}^2$.
Solid lines correspond to the extragalactic contribution, whereas dashed lines show the Galactic Center component.}
\label{fig:DifferentialRateComparison}
\end{figure}

In the case of isoscalar interactions,
$f^p=f^n=1$
and
$\sigma_p=\sigma_n$,
the above expressions simplify to
\begin{align}
    \sigma^{\rm coh}_{{\rm SI},{\cal T}}
    &=
    \sigma_p
    \left(
    \frac{\mu_{\cal T}}{\mu_p}
    \right)^2
    A_{\cal T}^2,
    \\
    \sigma^{\rm inc}_{{\rm SI},{\cal T}}
    &=
    \sigma_p A_{\cal T}.
\end{align}
The recoil spectrum induced by the BDM particles produced in cosmological semi-annihilations is readily obtained by inserting the differential BDM flux of Eq.~(\ref{eq:diff_flux}), evaluated at redshift zero, into Eq.~(\ref{eq:diff_rate}). In Fig.~\ref{fig:DifferentialRateComparison} we show, for illustration, the recoil spectrum for scattering on a $^{131}$Xe target assuming
$\langle \sigma_{2\to1} v\rangle
=
10^{-26}\,{\rm cm}^3\,{\rm s}^{-1}$,
$\sigma_p=10^{-31}\,{\rm cm}^2$,
and DM masses
$m_\chi=10$ MeV (violet),
$100$ MeV (red),
and $1$ GeV (green).
For comparison, we also display the spectra obtained in Ref.~\cite{BetancourtKamenetskaia:2025noa} for semi-annihilations occurring in the Galactic Center and for the same choice of parameters, which leads to a differential flux given by:
\begin{equation}\label{eq:diff_flux_gal}
    \frac{d\Phi_{\rm gal}}{dT_\chi}=\Phi_{\rm gal}\delta\left(T_\chi-\frac{m_\chi}{4}\right),\hspace{0.5cm}\Phi_{\rm gal}=3.2\times10^{-3}\mathrm{cm}^{-2}\mathrm{s}^{-1}\left(\frac{m_\chi}{100~\rm MeV}\right)^{-2}\left(\frac{\langle\sigma_{2\to1}v\rangle}{10^{-26}~\mathrm{cm}^3\mathrm{s}^{-1}}\right).
\end{equation}

It follows from the figure that the recoil rate generated by cosmological semi-annihilations is comparable to the Galactic Center contribution at low DM kinetic energy. Consequently, this contribution can increase the sensitivity of direct detection experiments to semi-annihilating DM scenarios. 

To estimate the sensitivity of current and future experiments, we compute the total event rate by integrating Eq.~\eqref{eq:diff_rate} over the recoil-energy window of the experiment and summing over all target nuclei:
\begin{equation}
\label{eq:int_scatt_rate}
    R
    =
    \sum_{\cal T}
    \int dT_{\cal T}\,
    \frac{dR_{\cal T}}{dT_{\cal T}}.
\end{equation}
The expected number of events, $N_{\rm exp}$, is then obtained by multiplying the total rate by the exposure.

\begin{figure}[t!]
\centering
\includegraphics[width=0.98\textwidth]{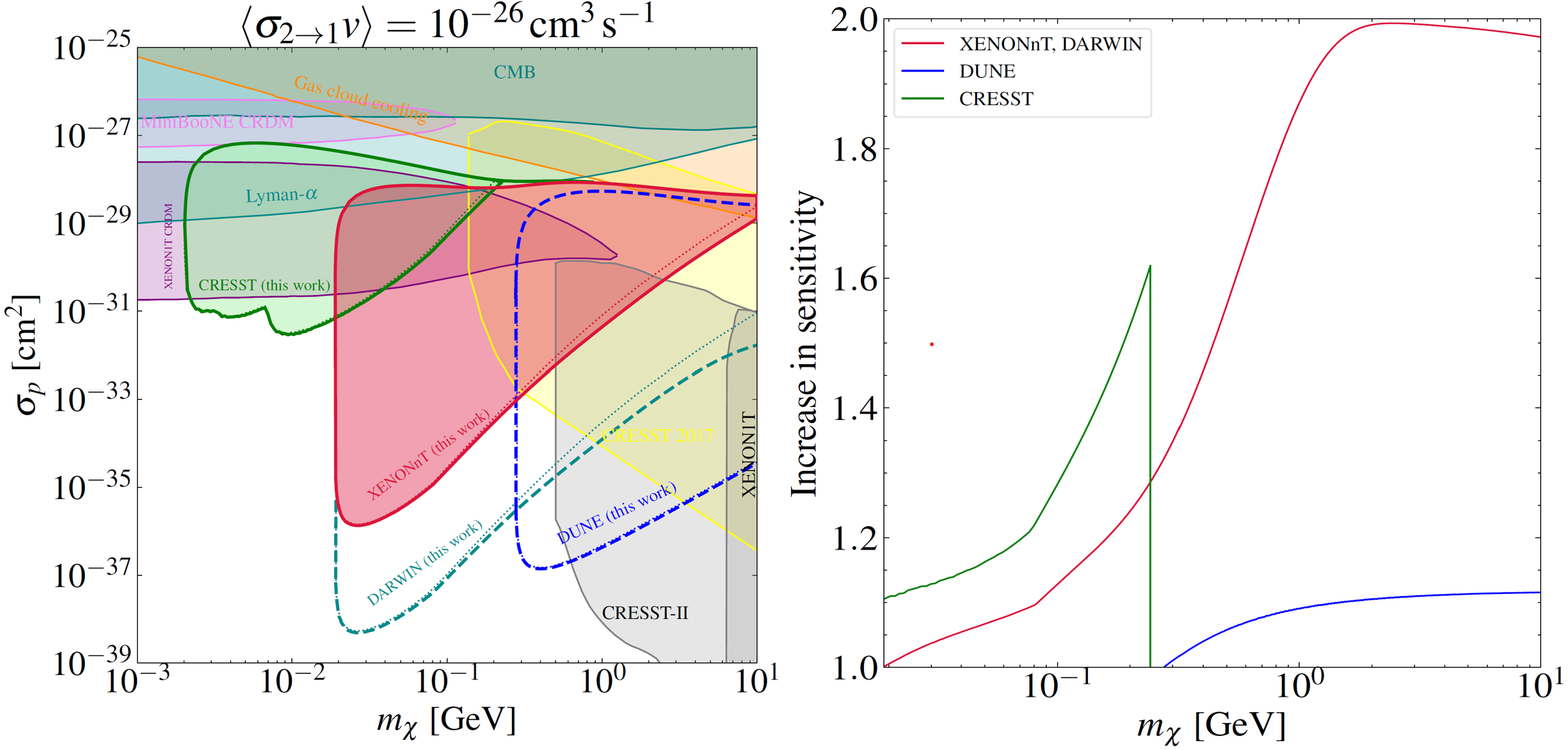}
\caption{\textit{Left panel:} Excluded regions for the spin-independent DM-proton cross section from XENONnT (red region), CRESST (green region), as well as the projected sensitivities for DARWIN (dashed dark cyan line) and DUNE (dashed blue line). We assume a cross section for the semi-annihilation
$\langle\sigma_{2\to1} v\rangle=10^{-26}~\mathrm{cm}^3~\mathrm{s}^{-1}$. The dotted lines correspond to the constraints obtained by considering only the Milky Way halo modeled with an NFW profile~\cite{BetancourtKamenetskaia:2025noa}. \textit{Right panel}: Increase in the experimental sensitivity of each experiment when the extragalactic component to the BDM flux is added to the galactic component. The abrupt end in the CRESST curve corresponds to the DM masses that cannot be probed due to the attenuation effect (see Ref.~\cite{BetancourtKamenetskaia:2025noa}).}
\label{fig:ExtragalacticConstraints}
\end{figure}

In the left panel of Fig.~\ref{fig:ExtragalacticConstraints} we present conservative upper limits on the spin-independent DM--proton scattering cross section as a function of the DM mass, derived from the non-observation of a signal in XENONnT~\cite{XENON:2023cxc} and CRESST-II~\cite{CRESST:2015txj}. The limits are obtained following the procedure of Ref.~\cite{BetancourtKamenetskaia:2025noa}, assuming a semi-annihilation cross section
$\langle\sigma_{2\to1} v\rangle
=
10^{-26}\,{\rm cm}^3\,{\rm s}^{-1}$. For comparison, we also show the constraints from the two CRESST phases~\cite{CRESST:2015txj,CRESST:2017ues}, XENON1T~\cite{XENON:2017vdw}, the BDM component generated through cosmic-ray upscattering~\cite{Bringmann:2018cvk}, as well as limits arising from the impact of DM--proton interactions on the cosmic microwave background~\cite{Xu:2018efh}, the Lyman-$\alpha$ forest~\cite{Rogers:2021byl}, and gas cloud cooling~\cite{Bhoonah:2018wmw}. We additionally show the projected sensitivities of the future DARWIN~\cite{DARWIN:2016hyl} and DUNE~\cite{DUNE:2015lol} experiments. The solid and dashed lines show the constraints when both the extragalactic (Eq.~\eqref{eq:diff_flux}) and galactic (Eq.~\eqref{eq:diff_flux_gal}) fluxes are considered, while the dotted lines indicate the corresponding constraints obtained when only the Milky Way halo contribution is included, assuming an NFW profile.

This figure demonstrates that the cosmological BDM component can enhance the sensitivity of direct detection experiments by a factor of ${\cal O}(1)$, compared to the calculation that only considers the galactic component.  For semi-annihilation cross sections close to the canonical thermal freeze-out value, CRESST excludes scattering cross sections as small as
$10^{-32}\,{\rm cm}^2$
for
$m_\chi\simeq10$ MeV,
while XENONnT reaches sensitivities down to
$10^{-34}\,{\rm cm}^2$
for
$m_\chi\simeq300$ MeV.
Future experiments will further extend this reach: DUNE could probe cross sections as small as
$10^{-37}\,{\rm cm}^2$,
while DARWIN may become sensitive to values around
$10^{-39}\,{\rm cm}^2$,
thereby entering the femtobarn regime for sub-GeV DM.

In the right panel of Fig.~\ref{fig:ExtragalacticConstraints} we quantify the improvement in experimental sensitivity brought by the extragalactic BDM flux. For each experiment we plot the ratio of the constraint obtained using only the galactic contribution to that obtained when both the galactic and extragalactic fluxes are included. As apparent from the plot, the extragalactic BDM component enhances slightly the reach of direct detection experiments, although it should be stressed that the modeling of this component is still subject to large uncertainties, and that the actual increase in sensitivity could be larger or smaller than the one we obtain.

\section{Conclusions}
\label{sec:conclusions}

We have considered the dark matter semi-annihilation scenario $\chi \chi \rightarrow \chi^c \phi$, with $\chi$ the dark matter particle and $\phi$ dark radiation, and 
we have analyzed the signals associated with the semi-annihilation in structures at high redshift, and specifically due to the production of a flux of boosted dark matter particles. Using a state-of-the-art model to calculate the enhancement in the annihilation rate at high redshift, we have found that the resulting flux is not negligible and could have implications for structure formation and (in some scenarios) for direct detection experiments. 

Requiring that the BDM number density does not represent more than 1\% of the total DM density,  we have derived a conservative model-independent limit on the semi-annihilation cross section, which reads $\langle\sigma_{2\to1} v\rangle\leq4.2\times10^{-19}~\left(m_\chi/1~\rm GeV\right)~\mathrm{cm}^3/{\rm s}$. For the scenario where dark radiation is identified with neutrinos, $\chi\chi\rightarrow \chi^c \nu$, these structure-formation bounds are typically weaker than those from neutrino telescopes. However, they become the dominant constraint for $m_\chi\lesssim2~\rm MeV$, where the neutrino energy falls below the threshold of production of charged particles via the charged current interaction.

We have also studied the impact of the cosmological BDM flux on direct detection experiments in scenarios where the DM interacts with the nucleus. Due to their large velocities, boosted particles can induce nuclear recoils at energies well above those expected from the standard halo component. We find that the contribution from cosmological semi-annihilations can be comparable to that from the Milky Way halo, thereby enhancing the sensitivity of direct detection experiments. Explicitly, for a thermal relic cross‑section $\langle\sigma_{2\to1}v\rangle = 10^{-26}\,\text{cm}^3/\text{s}$, our analysis shows that CRESST excludes DM–proton scattering cross sections as low as $10^{-32}\,\text{cm}^2$ for $m_\chi \simeq 10\,\text{MeV}$, while XENONnT reaches $10^{-34}\,\text{cm}^2$ for $m_\chi \simeq 300\,\text{MeV}$. Future experiments will extend this reach further: DUNE could probe cross sections down to $10^{-37}\,\text{cm}^2$, and DARWIN may reach values around $10^{-39}\,\text{cm}^2$, entering the femtobarn regime for sub‑GeV DM.

\section*{Acknowledgements}
We are grateful to Yidong Song for identifying a numerical error in the first version of the manuscript. This work is supported by the Collaborative Research Center SFB1258 and by the Deutsche Forschungsgemeinschaft (DFG, German Research Foundation) under Germany's Excellence Strategy - EXC-2094 - 390783311. The work of BBK is supported by IBS under the project code IBS-R018-D3. The work of MR is supported by the KIAS Individual Grant PG108401 at Korea Institute for Advanced Study.

\printbibliography

\end{document}